\documentclass[sigplan,screen]{acmart}

\AtBeginDocument{%
  \providecommand\BibTeX{{%
    \normalfont B\kern-0.5em{\scshape i\kern-0.25em b}\kern-0.8em\TeX}}}

\newcommand{\NPM}{NPM}
\newcommand{\NodeJS}{NodeJS}

\newcommand{\CountTotalModulesDefinitelyTyped}{6029}
\newcommand{\CountModulesWithRepositoryUrl}{4974}

\newcommand{\CountModulesWithCodeExamples}{2260}
\newcommand{\CountModulesWorkingCodeExamples}{946}
\newcommand{\CountModulesRunTimeInfoExtracted}{436}
\newcommand{\CountModulesGeneratedDeclarationFile}{249}
\newcommand{\CountModulesOnlySolvableDifferences}{111}

\newcommand{\CountModulesWithoutReadmeFile}{682} 
\newcommand{\CountModulesNotWorkingCodeExamples}{1314} 
\newcommand{\CountModulesRunTimeInfoCouldNotBeExtracted}{510} 
\newcommand{\CountModulesDeclarationFileCouldNotBeGenerated}{138} 
\newcommand{\secref}[1]{Section~\ref{#1}}
\newcommand{\figref}[1]{Figure~\ref{#1}}
\newcommand{\coderef}[1]{Listing~\ref{#1}}

\usepackage{listings}

\usepackage{cleveref}
\usepackage{pgfplots}
\usepackage{tikz}
\usetikzlibrary{positioning}

\usepackage{xcolor}
\usepackage{tcolorbox}
\tcbuselibrary{breakable}

\usepackage{subcaption}

\usepackage[htt]{hyphenat}

\usepackage[utf8]{inputenc}
\usepackage[T1]{fontenc}
\usepackage{microtype}

\definecolor{backgroundgray}{rgb}{.94,.94,.94}

\lstdefinelanguage{JavaScript}{
  keywords={typeof, new, true, false, catch, function, return, null, catch, switch, var, if, in, while, do, else, case, break, undefined,const},
  keywordstyle=\color{black}\bfseries,
  ndkeywords={class, export, boolean, throw, implements, import, this},
  ndkeywordstyle=\color{black}\bfseries,
  identifierstyle=\color{darkgray},
  sensitive=false,
  comment=[l]{//},
  morecomment=[s]{/*}{*/},
  commentstyle=\color{lightgray}\slshape,
  stringstyle=\color{darkgray}\ttfamily,
  morestring=[b]',
  morestring=[b]"
}

\lstdefinelanguage{TypeScript}{
  keywords={typeof, new, true, false, catch, function, return, null, catch, switch, var, if, in, while, do, else, case, break, undefined,const},
  keywordstyle=\color{black}\bfseries,
  ndkeywords={class, export, boolean, throw, implements, import, this, declare, constructor, namespace, interface},
  ndkeywordstyle=\color{black}\bfseries,
  identifierstyle=\color{darkgray},
  sensitive=false,
  comment=[l]{//},
  morecomment=[s]{/*}{*/},
  commentstyle=\color{lightgray}\slshape,
  stringstyle=\color{darkgray}\ttfamily,
  morestring=[b]',
  morestring=[b]"
}

\graphicspath{{./graphics/}}

\copyrightyear{2021}
\acmYear{2021}
\setcopyright{acmlicensed}\acmConference[MPLR '21]{Proceedings of the 18th ACM SIGPLAN International Conference on Managed Programming Languages and Runtimes}{September 29--30, 2021}{Münster, Germany}
\acmBooktitle{Proceedings of the 18th ACM SIGPLAN International Conference on Managed Programming Languages and Runtimes (MPLR '21), September 29--30, 2021, Münster, Germany}
\acmPrice{15.00}
\acmDOI{10.1145/3475738.3480941}
\acmISBN{978-1-4503-8675-3/21/09}

\begin{document}

\title{Generation of TypeScript Declaration Files from JavaScript Code}


\author{Fernando Cristiani}
\affiliation{%
  \institution{Hochschule Karlsruhe}
  \city{Karlsruhe}
  \country{Germany}}
\email{fernando.cristiani@hs-karlsruhe.de}

\author{Peter Thiemann}
\affiliation{%
  \institution{Albert-Ludwigs-Universität Freiburg}
  \city{Freiburg}
  \country{Germany}
}
\email{thiemann@informatik.uni-freiburg.de}


\begin{abstract}
  Developers are starting to write large and complex applications in
  TypeScript, a typed dialect of JavaScript. TypeScript applications
  integrate JavaScript libraries via typed descriptions of their APIs
  called declaration files. DefinitelyTyped is the standard public
  repository for these files.
  The repository is populated and maintained manually by volunteers, which
  is error-prone and time consuming. Discrepancies between a
  declaration file and the JavaScript implementation lead to
  incorrect feedback from the TypeScript IDE and, thus, to incorrect uses
  of the underlying JavaScript library.

  This work presents \texttt{dts-generate}, a tool that generates
  TypeScript declaration files for JavaScript libraries uploaded to the \NPM{}
  registry. It extracts code examples from the documentation written by
  the developer, executes the library driven by the examples, gathers
  run-time information, and generates a declaration file based on this
  information. To evaluate the tool, \CountModulesGeneratedDeclarationFile{} declaration files were
  generated directly from an \NPM{} module and \CountModulesOnlySolvableDifferences{} of these were
  compared with the corresponding declaration file provided on
  DefinitelyTyped.  All these files either
  exhibited no differences at all or differences that can be resolved by
  extending the developer-provided examples.
\end{abstract}




\maketitle

\section{Introduction}
\label{sec:introduction}
JavaScript is the most popular language for writing web
applications \cite{github-statistics}. It is also increasingly used
for back-end applications running in \NodeJS{}, a JavaScript-based
server-side platform. JavaScript is appealing to developers because
its forgiving dynamic typing enables 
them to create simple pieces of code very quickly and proceed on a
trial-and-error basis.

JavaScript was never intended to be more than a
scripting language and, thus, lacks features for maintaining and evolving large
codebases. However, nowadays developers create large and complex
applications in JavaScript. 
Mistakes such as mistyped property
names and misunderstood or unexpected type coercions cause developers
to spend a significant amount of time in debugging. There is ample
evidence for such mishaps. For example, a 
JavaScript code blog\footnote{\url{https://wtfjs.com}} collects experiences
from developers facing unexpected situations while programming in
JavaScript. \coderef{code:introduction-javascript-wtfs} exposes some
of these unintuitive JavaScript behaviors. 

\begin{lstlisting}[
  caption={\textbf{Unintuitive JavaScript behavior:} Falsy values,
  \lstinline!null! vs. \lstinline!undefined!,
\lstinline!typeof!, and type coercion},
  language=JavaScript,
label=code:introduction-javascript-wtfs,
  float=bp,
  captionpos=b
]
"0" == false; // true
true == "1.00"; // true
false == "    \n\r\t     "; // true
false == []; // true
0 == []; // true
null == undefined; // true
null + undefined + [1, 2, 3] // 'NaN1,2,3' 
typeof null; // object
null instanceof Object; // false
[1] + 1; //'11'
[2] == "2"; // true
"hello world".lenth + 1 // NaN | note 'lenth' instead of 'length'
[1, 15, 20, 100].sort() // [ 1, 100, 15, 20 ]
\end{lstlisting}

The cognitive load produced by these behaviors is mitigated in languages that use
build tools based on type information. This insight motivated the
creation of TypeScript, a superset of JavaScript with expressive type
annotations \cite{typescript}. It has become a widely used alternative
among JavaScript developers, because it incorporates features that are
helpful for developing and maintaining large applications
\cite{DBLP:conf/icse/GaoBB17}. TypeScript enables the early detection
of several kinds of run-time errors and the integration of code intelligence
tools like autocompletion in an IDE.

Despite the advantages, it is unrealistic to expect the world to
switch to TypeScript in a day. Therefore, 
existing JavaScript libraries can be used in a TypeScript project by
adding a declaration file that describes the library's
API in terms of types. 
The DefinitelyTyped repository~\cite{definitely-typed-repository} has
been created as a community effort to collect declaration files for
popular JavaScript libraries. At the time of writing it contains
declaration files for more than 6000 libraries.

Unfortunately, creation and maintenance of most declaration
files in DefinitelyTyped is conducted manually,
which is time consuming and  error-prone. Errors are aggravated because TypeScript takes a
declaration file at face value. Discrepancies between declaration and
implementing JavaScript library are not detected at 
compile time and result in incorrect behavior (e.g., autocompletion
hints) of the IDE. 
As TypeScript does not perform any run-time 
checking of types, these  discrepancies can lead to unexpected 
behavior and crashes. This kind of experience can lead to longish debugging sessions,  developer
frustration, and decreasing confidence in the tool chain.

\subsection{Approach}
\label{sec:approach}

We aim to improve on this situation by providing a tool that generates
TypeScript declaration files automatically. To do so, we rely on run-time information
gathered from the examples provided
by the developer as part of the documentation of a library. If these
examples are sufficiently rich, then our toolchain can generate
high-quality declaration files for the library. This way we take
advantage of 
best practices in the dynamic languages community which favors
examples and tests over writing out type signatures.

As an example, consider the NPM\footnote{NPM is the \href{https://www.npmjs.com/}{Node package
  manager}, an online repository for JavaScript modules.} module
\texttt{abs} that
``computes the absolute path of an input''. Its documentation contains
the following three
examples\footnote{\url{https://www.npmjs.com/package/abs}}. 

\begin{lstlisting}[language=JavaScript,numbers=none]
const abs = require("abs");
 
console.log(abs("/foo"));
// => "/foo"
 
console.log(abs("foo"));
// => "/path/to/where/you/are/foo"
 
console.log(abs("~/foo"));
// => "/home/username/foo"
\end{lstlisting}
From the examples, our tool generates the following TypeScript declaration file.
\begin{lstlisting}[language=TypeScript,numbers=none]
export = Abs;
declare function Abs(input: string): string;
\end{lstlisting}
It used to be equivalent to the declaration file provided in the
DefinitelyTyped
repository\footnote{\url{https://github.com/DefinitelyTyped/DefinitelyTyped/blob/master/types/abs/index.d.ts}},
but the maintainer indicates (since May 10, 2021) that the parameter is optional:
\begin{lstlisting}[language=TypeScript,numbers=none]
declare function abs(input?: string): string;
export = abs;
\end{lstlisting}
This case is quite frequent. The information in our declaration file is
correct, but incomplete, because the developers did not document the
fact that the input is optional. If the developer had provided the
call  \lstinline/abs()/ as an additional 
example, then our tool would have generated exactly
the declaration on DefinitelyTyped! We call this case a solvable
difference between our tool's output and the DefinitelyTyped
declaration file.


Our tool \texttt{dts-generate} is a first step to explore the possibilities for
generating useful declaration files from run-time information.

\texttt{dts-generate} comes with a framework that 
supports the generation of declaration files for an existing
JavaScript library published to the \NPM{} registry. The tool gathers
data flow and type information at run time to generate a declaration
file based on that information.

The main contribution of our tool is twofold:
\begin{enumerate}
\item
  We do not rely on static analysis, which is hard to implement
  soundly and precisely. It is also prone to maintenance problems
  when keeping up with JavaScript's frequent language updates.
\item
  We extract example code from the programmer's library
  documentation and rely on dynamic analysis to extract typed usage
  patterns for the library from the example runs.
\end{enumerate}


Our implementation of code instrumentation  to gather data flow
information and type information at run time is based on
Jalangi \cite{DBLP:conf/sigsoft/SenKBG13}, a configurable
framework for dynamic analysis for JavaScript.
The implementation and the results are available in the following repositories.

\begin{itemize}\footnotesize
  \item \url{https://github.com/proglang/run-time-information-gathering/tree/v1.1.0}
  \item \url{https://github.com/proglang/ts-declaration-file-generator/tree/v1.5.0}
  \item \url{https://github.com/proglang/ts-declaration-file-generator-service/tree/v1.2}
  \item \url{https://github.com/proglang/dts-generate-method/tree/v1.3.1}
  \item \url{https://github.com/proglang/dts-generate-results/tree/v1.4.0}
\end{itemize}

\subsection{Contributions}
\label{sec:contributions}

\begin{itemize}
\item A framework that extracts code examples from the
  documentation of an \NPM{} package and collects run-time type
  information from running these examples (Sections~\ref{sec:initial-code-base}
  and~\ref{sec:run-time-information}). 

\item Design and implementation of the tool \texttt{dts-generate}, a command line
  application that generates a valid TypeScript declaration file for
  an \NPM{} package using run-time information
  (Section~\ref{sec:typescr-decl-file}). 

\item A comparator for TypeScript declaration files
  (Section~\ref{sec:dts-compare}) for evaluating our framework and to
  detect   incompatibilities when evolving JavaScript modules.
\item An evaluation of our framework (Sections~\ref{sec:dts-generate-evaluation}
  and~\ref{sec:results}). We examined all \CountTotalModulesDefinitelyTyped{} entries in 
  the DefinitelyTyped repository and found \CountModulesGeneratedDeclarationFile{} sufficiently
  well documented \NPM{} packages, on which we ran \texttt{dts-generate}
  and compared the outcome with the respective declaration file from the
  DefinitelyTyped repository. 
\end{itemize}

\section{Motivating Example}
\label{sec:motivating-example}
\begin{figure}[tp]
  \centering
    \begin{lstlisting}[language=TypeScript,numbers=none]
export = GlobToRegexp;

declare function GlobToRegexp(
    glob: string, 
    opts?: GlobToRegexp.I__opts): RegExp;
declare namespace GlobToRegexp {
  export interface I__opts {
    'extended'?: boolean;
    'globstar'?: boolean;
    'flags'?: string;
  }

}
    \end{lstlisting}
  \begin{center}
    \includegraphics[width=\linewidth]{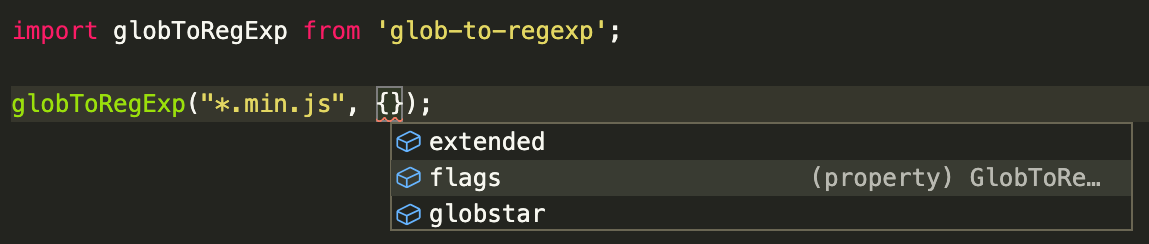}
  \end{center}

  \caption{Generated declaration file for \texttt{glob-to-regexp}} 
  \label{fig:motivating-example-glob-to-regexp-vscode}
\end{figure}

The \NPM{} package \texttt{glob-to-regexp} provides functionality to turn a
glob expression for matching filenames in the shell into a regular
expression\footnote{\url{https://www.npmjs.com/package/glob-to-regexp}}. It 
has about 9.9M weekly downloads and 188 \NPM{} packages depend on it. If
a developer creates or extends TypeScript code that depends on the
\texttt{glob-to-regexp} library, the TypeScript
compiler and IDE require a declaration file for that library to
perform static checking and code completion, respectively. With
\texttt{dts-generate} we automatically generate a TypeScript
declaration file for \texttt{glob-to-regexp}. Our tool downloads the \NPM{} 
package, runs the examples extracted from its documentation, gathers
run-time information, and generates a TypeScript declaration
file. As the package is insufficiently documented, we provide one
additional example to generate the result 
shown in \figref{fig:motivating-example-glob-to-regexp-vscode} (see
discussion in Section~\ref{sec:typescr-decl-file}).  The interface is
detected correctly. Optional parameters are detected. This declaration is
ready for use in a TypeScript project:  the generated file works
properly with Visual Studio Code\footnote{\url{https://code.visualstudio.com}}. If the
\texttt{glob-to-regexp} package gets modified in the future, a new declaration
file can be generated automatically using
\texttt{dts-generate}. Our comparator tool (Section~\ref{sec:dts-compare}) can compare the new file
for incompatibilities with the previous declaration file.

\section{TypeScript Declaration Files}
\label{sec:typescr-decl-files}

The generated declaration file shown in \figref{fig:motivating-example-glob-to-regexp-vscode}
describes a package with a single exported function. The module name
is derived from the NPM module name by camlization\footnote{i.e.,
  transforming to camel case.}. The first parameter is of type
\texttt{string} and the second one is an optional object described by
an interface. To avoid name clashes, \texttt{dts-generate} creates a
namespace that corresponds to the module name. This pattern is a best
practice to  organize types declared in a declaration file
\cite{typescript-namespaces}. In the example, the interface
\lstinline[language=TypeScript]{I__opts} is declared in namespace
\lstinline[language=TypeScript]{GlobToRegexp}. Hence, its uses
must be qualified by the namespace as in
\lstinline[language=TypeScript]{GlobToRegexp.I__opts}. The name of the
interface is derived from the name of the formal parameter. 

This file is an instance of one of the standard templates for writing
declaration
files\footnote{\url{https://www.typescriptlang.org/docs/handbook/declaration-files/templates.html}}:
\textbf{module}, \textbf{module-class}, and
\textbf{module-function}. 
Each template corresponds to a different way of organizing the exports
of a JavaScript library. The choice of the template depends on the
structure of the underlying JavaScript library:
\begin{description}
\item[module] several exported functions,
\item[module-class] a class-like structure,
\item[module-function] exactly one exported function.
\end{description}
There are further templates, but most libraries fall in one of these
three categories.
Both libraries \texttt{glob-to-regexp} and \texttt{abs} are instances of the
\textbf{module-function} template.

\begin{lstlisting}[
  caption={Example for \textbf{module-function} template},
  language=bash,
label=code:dts-generate-example,
  float,captionpos=b
]
$ ./dts-generate abs
$ cat output/abs/index.d.ts 
export = Abs;

declare function Abs(input: string): string;
\end{lstlisting}

\begin{figure*}[t]
\centering
\begin{subfigure}[t]{0.48\linewidth}
  \begin{lstlisting}[language=JavaScript,numbers=none]
var greet = require("./greet-settings-module");

greet({
greeting: "hello world",
duration: 4000
});

greet({
greeting: "hello world",
color: "#00ff00"
});
  \end{lstlisting}
  \caption{Example for \textbf{module-function} template}
  \label{fig:example-module-function}
\end{subfigure}
\hfill
\begin{subfigure}[t]{0.48\linewidth}
  \begin{lstlisting}[language=TypeScript,numbers=none]
export = GreetSettingsModule;

declare function GreetSettingsModule(settings: GreetSettingsModule.I__settings): void;
declare namespace GreetSettingsModule {
export interface I__settings {
  'greeting': string;
  'duration'?: number;
  'color'?: string;
}

}
  \end{lstlisting}
  \caption{Declaration file for \textbf{module-function} template}
  \label{fig:template-module-function}
\end{subfigure}

\begin{subfigure}[t]{0.48\linewidth}
    \begin{lstlisting}[language=JavaScript,numbers=none]
var myLib = require("./greet-module");

var result = myLib.makeGreeting("hello, world");
console.log("The computed greeting is: " + result);

var goodbye = myLib.makeGoodBye();
console.log("The computed goodbye is: " + goodbye);    
    \end{lstlisting}
  \caption{Example for \textbf{module} template}
  \label{fig:example-module}
  \end{subfigure}
  \hfill
  \begin{subfigure}[t]{0.48\linewidth}
    \begin{lstlisting}[language=TypeScript,numbers=none]
export function makeGreeting(str: string): string;
export function makeGoodBye(): string;        
    \end{lstlisting}
    \caption{Declaration file for \textbf{module} template}
    \label{fig:template-module}
  \end{subfigure}

  \begin{subfigure}[t]{0.48\linewidth}
    \begin{lstlisting}[language=JavaScript,numbers=none]
var Greeter = require("./greet-classes-module.js");

var myGreeter = new Greeter("hello, world");
myGreeter.greeting = "howdy";
myGreeter.showGreeting();
    \end{lstlisting}
    \caption{Example for \textbf{module-class} template}
    \label{fig:example-class}
  \end{subfigure}
  \hfill
  \begin{subfigure}[t]{0.48\linewidth}
    \begin{lstlisting}[language=TypeScript,numbers=none]
export = Greeter;

declare class Greeter {
constructor(message: string);
showGreeting(): void;
}

declare namespace Greeter {
}
    \end{lstlisting}
    \caption{Declaration file for \textbf{module-class} template}
    \label{fig:template-class}
  \end{subfigure}

\caption{Example uses and declaration files for different templates}
\label{fig:typescript-templates-by-example}
\end{figure*}

The TypeScript project provides a guide on how to write high-quality declaration
files\footnote{\url{https://www.typescriptlang.org/docs/handbook/declaration-files/by-example.html}}. The guide
explains the main concepts through examples. We selected one example for each template
from the guide and
used \texttt{dts-generate} to generate a corresponding declaration file, as shown in
\figref{fig:typescript-templates-by-example}.

The Typescript project provides a package \texttt{dts-gen}, which
generates a template for a declaration file by analyzing the structure
of the module. The resulting file is intended as a starting point for further
manual development of the declaration file. 
Unlike \texttt{dts-gen} we determine the kind of declaration file by
examining the \emph{usage} of the module. If the imported entity is
used solely as a function as in
Figure~\ref{fig:example-module-function}, the 
generated declaration file follows the \textbf{module-function}
template as shown in Figure~\ref{fig:template-module-function}. 
If the example only accesses properties of the imported entity as in
Figure~\ref{fig:example-module}, we generate the declarations
according to the \textbf{module} template
(Figure~\ref{fig:template-module}). 
If the imported entity is used with \lstinline/new/ to create new instances as in
Figure~\ref{fig:example-class}, we generate a \textbf{module-class} template (Figure~\ref{fig:template-class}).
While it is possible to create a library of undetermined category, our selection is driven
by the examples in the documentation and 
hence reflects the developer's intent.











\section{The Generation of TypeScript Declaration Files}
\label{sec:gener-typescr-decl}
\begin{figure}[tp]
  \centering
  \includegraphics[width=1\linewidth]{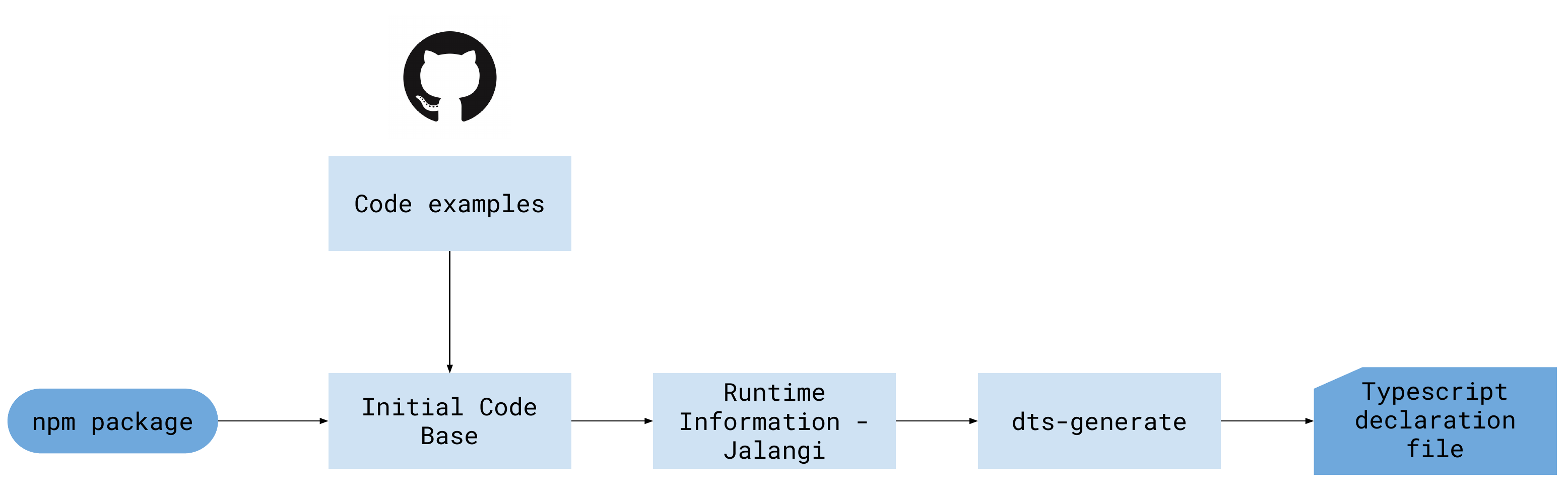}
  \caption[dts-generate - Architecture overview]{dts-generate - Architecture
    overview
  } 
  \label{fig:tsd_generation_method_block_diagram}
\end{figure}

This section gives an overview of our approach to generating
TypeScript declaration files from a JavaScript library packaged in
\NPM. \figref{fig:tsd_generation_method_block_diagram} gives a rough
picture of the inner working of our tool \texttt{dts-generate}. The
input is an \NPM{} package and the output is a TypeScript declaration
file for the package if it is ``sufficiently documented'', which we
substantiate in the next subsection.


As \texttt{dts-generate} is based on run-time information, we need code
fragments that import and use the JavaScript library and we need to be
able to obtain this run-time information while running the code.
We explain in Section~\ref{sec:initial-code-base} how we obtain code
fragments. 
To obtain run-time information, we instrument
the code fragments and the entire code base of the library with
Jalangi2 \cite{DBLP:conf/sigsoft/SenKBG13}. We extend Jalangi's
analysis modules to gather data flow and type information as required
for our purposes and explained in
Section~\ref{sec:run-time-information}.
\texttt{dts-generate} supports full ECMAScript~5.1, which is
the latest JavaScript version supported by Jalangi2. 

A second independent block explained in
Section~\ref{sec:typescr-decl-file} uses the run-time information to
generate a TypeScript declaration file. This part infers the overall
structure of the JavaScript library, its interfaces, and the types
from the run-time information.  The resulting declaration file is
ready for use in the development process. Its contents mimic the usage
of the library in the example code fragments and match the structure
of the JavaScript library under analysis, so that the JavaScript code
generated after compiling the TypeScript code runs without
interface-related errors.

The command-line interface is very simple and inspired by the \texttt{dts-gen} tool
\cite{dts-gen} (see \coderef{code:dts-generate-example}). The only required
argument is the name of the module published to the \NPM{} registry. 

\subsection{Initial Code Base}
\label{sec:initial-code-base}


There are several options to obtain code fragments that exercise the
library code in a meaningful way. In order to exercise library $A$,
one might
\begin{enumerate}
\item\label{item:1} execute code that imports library $A$;
\item\label{item:2} execute the test cases that come with library $A$;
\item\label{item:3} execute code fragments extracted from library $A$'s documentation.
\end{enumerate}

Option~\ref{item:1} requires finding a library $B$ that depends on
$A$. While this information is straightforward to obtain from \NPM, it
is potentially costly to download and instrument $B$ for use with
Jalangi. Moreover, this option delegates our problem
to library $B$, which also needs code to drive it. 

We considered option~\ref{item:2} under the assumption that most
libraries would come with test cases. However, there is no standard
for testing JavaScript code so that test cases were difficult to reap
from the \NPM{} packages: they use different directory structures,
employ differing (or no) testing tools, or do not have tests
at all. Moreover, developers try to cover corner cases or to trigger errors by using illegal
input values in their tests. Such tests are not really useful for describing a library
from the consumer's point of view. For example, a developer might write a test invoking a
function with \texttt{null} just to validate that it throws an error
in such a scenario. 

In the end, option~\ref{item:3} was the most viable option to gather run-time information
even though there is no standard for documentation, either. However, almost all repositories
contain README files where the library authors briefly describe in prose what
the code does, which problem it solves, how to install the
application, how to build the code, etc. It is very common that
developers provide code examples in the README files to show how the
library works and how to use it. These code fragments showcase common use cases rather
than stress-testing the implementation (a problem of
option~\ref{item:2}) because they are meant to be instructive to users of the package.

Obtaining code examples for a specific \NPM{} package is done in three steps.
\begin{itemize}
\item Obtain the  repository's URL with the command
  \texttt{npm view <PACKAGE> repository.url}

\item Retrieve the README file from the top-level directory of the repository.

\item Extract the code examples from the README file. To this end,
  observe that README files are
  written using Markdown\footnote{\url{https://www.markdownguide.org}}, a
  popular markup language, where most
  code examples are presented in code blocks labeled with the programming
  language. Hence, we
  retrieve the code examples from code blocks labeled \texttt{js} or
  \texttt{javascript}, which both stand for JavaScript.
\end{itemize}

\coderef{code:code-example-extracted-motivating-example} shows
the examples extracted from  the README file of the \texttt{glob-to-regexp} package in
that way.

\begin{lstlisting}[
  caption={Code extracted from the \texttt{glob-to-regexp} package},
  language=JavaScript,
  label=code:code-example-extracted-motivating-example,
  float,captionpos=b
]
var globToRegExp = require('glob-to-regexp');
var re = globToRegExp("p*uck");
re.test("pot luck"); // true
re.test("pluck"); // true
re.test("puck"); // true

re = globToRegExp("*.min.js");
re.test("http://example.com/jquery.min.js"); // true
re.test("http://example.com/jquery.min.js.map"); // false

re = globToRegExp("*/www/*.js");
re.test("http://example.com/www/app.js"); // true
re.test("http://example.com/www/lib/factory-proxy-model-observer.js"); // true

// Extended globs
re = globToRegExp("*/www/{*.js,*.html}", { extended: true });
re.test("http://example.com/www/app.js"); // true
re.test("http://example.com/www/index.html"); // true
\end{lstlisting}

Obtaining the code fragments from the examples provided in the
README files of the repository proved to be an
appropriate and pragmatic way of extracting the developer's
intention. It provides an initial code base with meaningful
examples.

\subsection{Run-time Information Gathering}
\label{sec:run-time-information}
The run-time information block in
\figref{fig:tsd_generation_method_block_diagram} gathers
type-related information as well as usage-related information for each function parameter.
Here are some examples.

\begin{itemize}
  \item Function \lstinline{f} was invoked where parameter
    \lstinline{a} held a value of type \lstinline{string} and
    \lstinline{b} a value of type \lstinline{number}. 
  \item Function \lstinline{foo} was invoked. Its parameter
    \lstinline{hello} was accessed as an object within the
    function. Its property \lstinline{world} was found to have a
    string value (see Figure~\ref{fig:example-get-field-interaction}).
  \item Function \lstinline{foo} was invoked. Its parameter
    \lstinline{bar} was accessed as an object. Its method
    \lstinline{hello} was invoked. The return value of this call was
    accessed as an object with property \lstinline{world}, which
    returned a string (see Figure~\ref{fig:example-method-call-interaction}).
  \item Parameter \lstinline{a} of function \lstinline{f} was used as
    operand for \lstinline{==}. 
\end{itemize}

Jalangi's configurable analysis modules enable
programming custom callbacks that can get triggered with virtually any
JavaScript event. Our instrumentation observes the following events: 
\begin{itemize}
  \item binary operations, like \lstinline{==}, \lstinline{+}, or
    \lstinline{===};
  \item variable declarations;
  \item function, method, or constructor invocations;
  \item access to an object's property;
  \item unary operations, like \lstinline{!} or \lstinline{typeof}.
\end{itemize}

The implementation stores these observations as entities called
\texttt{interactions}, which document each operation applied. They are used for translating, modifying, and
aggregating Jalangi's raw event information to get an application-specific data representation. Each function invocation is stored as
a \texttt{FunctionContainer} which contains an \texttt{ArgumentContainer} for each argument. Each \texttt{ArgumentContainer} contains 
the name and index of the argument and a collection of
\texttt{interactions} on the argument. As shown in \figref{fig:example-get-field-interaction}, the interaction \texttt{getField} gets triggered whenever 
a property of an object gets accessed. It tracks the name of the
accessed field and the type of the returned value. The invocation of a
method is tracked with the \texttt{methodCall} interaction. An
interaction can have a \texttt{followingInteractions} property if
there are further operations on the return value of the interaction. It is used for inferring nested interfaces.
That is, the \texttt{followingInteractions} recursively record \texttt{interactions} on the return value of \texttt{getField} or 
\texttt{methodCall}, taking care of object identities to avoid
looping. \figref{fig:example-method-call-interaction} provides an
example for both the \texttt{methodCall} interaction 
and the \texttt{followingInteractions} property. There is a
\texttt{usedAsArgument} interaction (not shown) that
gets triggered when an argument is used in the invocation of another function.

\begin{figure}[t]
  \centering
    \begin{lstlisting}[language=JavaScript,numbers=none]
function foo(hello) {
  if (hello.world) {
    // ...
} }

foo({ world: "..." });
    \end{lstlisting}
    \begin{lstlisting}[language=JavaScript,numbers=none,escapechar=!]
{ functionId_1: {
    //...
    functionName: 'foo',
    args: {
      0: {
        //...
        argumentName: "hello",
        interactions: [
          // ...
          !\textbf{\{}!
            !\textbf{code: "getField",}!
            !\textbf{field: "world",}!
            !\textbf{followingInteractions: [],}!
            !\textbf{returnTypeOf: "string",}!
          !\textbf{\},}!],},},},}
    \end{lstlisting}
  \caption{
      Example code for \texttt{getField} interaction
    }
  \label{fig:example-get-field-interaction}
\end{figure}

\begin{figure}[t]
      \begin{lstlisting}[language=JavaScript,numbers=none]
function foo(bar) {
  if (bar.hello().world) {
    // ...
  }
}

foo({
  hello: function () {
    return {
      world: "xxx",
};},});    
      \end{lstlisting}

      \begin{lstlisting}[language=JavaScript,numbers=none,escapechar=!]
{ functionId_1: {
    //...
    functionName: "foo",
    args: {
      0: {
        // ...
        argumentName: "bar",
        interactions: [
          // ...
          !\textbf{\{}!
            !\textbf{code: "methodCall",}!
            !\textbf{methodName: "hello",}!
            !\textbf{functionId: "functionId\_2",}!
            !\textbf{followingInteractions: [}!
              !\textbf{\{}!
                !\textbf{code: "getField",}!
                !\textbf{field: "world",}!
                !\textbf{followingInteractions: [],}!
                !\textbf{returnTypeOf: "string",}!
              !\textbf{\},}! !\textbf{],}! !\textbf{\},}!],},},},}
      \end{lstlisting}
  \caption{\texttt{}{methodCall} and \texttt{followingInteractions}}
  \label{fig:example-method-call-interaction}
\end{figure}

\begin{figure*}[t]
  \begin{subfigure}[t]{0.48\linewidth}
    \begin{lstlisting}[language=JavaScript,numbers=none]
const foo = require('foo');

foo.doSomething();
    \end{lstlisting}
  \caption{Example code for \texttt{module} template}
  \end{subfigure}
  \hfill
  \begin{subfigure}[t]{0.48\linewidth}
    \begin{lstlisting}[language=JavaScript,numbers=none,escapechar=!]
{
  functionId_1: {
    functionName: 'doSomething',
    //...
    !\textbf{isExported: false,}!
    !\textbf{requiredModule: 'foo'}!
  }
}    
    \end{lstlisting}
    \caption{Run-time information fragment for \texttt{module} template}
  \end{subfigure}

  \begin{subfigure}[t]{0.48\linewidth}
    \begin{lstlisting}[language=JavaScript,numbers=none]
const foo = require('foo');

foo();
    \end{lstlisting}
    \caption{Example code for \texttt{module-function} template}
  \end{subfigure}
  \hfill
  \begin{subfigure}[t]{0.48\linewidth}
    \begin{lstlisting}[language=JavaScript,numbers=none,escapechar=!]
{
  functionId_1: {
    functionName: 'foo',
    //...
    !\textbf{isExported: true,}!
    !\textbf{requiredModule: 'foo'}!
  }
}    
    \end{lstlisting}
    \caption{Run-time information for \texttt{module-function} template}
  \end{subfigure}
  \caption{Example code for properties \texttt{isExported} and \texttt{requiredModule} - Property \texttt{isExported} is
  \texttt{false} for \texttt{module} and \texttt{true} for \texttt{module-function}. Additionally, \texttt{requiredModule}
  is \texttt{foo} for both cases.
  }
  \label{fig:example-is-exported-required-module}
\end{figure*}

To obtain this information we wrap each function's argument in a wrapper object, which stores meta-information and provides the mapping of an observation 
to the corresponding \texttt{ArgumentContainer} or
\texttt{interaction}. For operators that are aware of object identity, such as 
\texttt{===} or \texttt{typeof}, we use the original values because
they could return different results for wrapper objects.

We also gather information to determine which  module template is most
appropriate. To this end
the property \texttt{requiredModule} stores the name of the module that declared the invoked function. If a function is explicitly 
exported by the module, the property \texttt{isExported} is set to \texttt{true} when it is invoked. If a function of an 
exported object is invoked, \texttt{isExported} will be \texttt{false} and \texttt{requiredModule} will contain the name of 
the required module, as shown in
\figref{fig:example-is-exported-required-module}. If a function is
used as a constructor, we set its \texttt{isConstructor} flag.


\subsection{TypeScript Declaration File Generation}
\label{sec:typescr-decl-file}

The next step after gathering the run-time information is
the generation of the declaration file (cf.\
\figref{fig:tsd_generation_method_block_diagram}).
%
Contrary to \texttt{dts-gen} which analyzes the shape of the exported module, we choose the
template based on the way the module is used. We analyze the properties \texttt{isExported},
\texttt{requiredModule}, and \texttt{isConstructor} from the
run-time information to distinguish
between \textbf{module-class} and \textbf{module-function}. If a function is invoked and
it is imported from the module we are analyzing we infer the template
\textbf{module-class} or \textbf{module-function}. We choose \textbf{module-class} if the
function is used as a constructor. Otherwise, we choose
\textbf{module-function}.
If no function is invoked, we use the \textbf{module}
template.

Our goal is to support the  most commonly used TypeScript
features first in our implementation. To this end,
the selection of typing features covered in the tool is driven by
an analysis of the 6648 declaration files in
the DefinitelyTyped repository. \footnote{As of commit
  \texttt{da93d7b13094bacb170ead3f4a289f3b8687e4e5} (April 4, 2020).}
We parsed each declaration file with 
\texttt{dts-parse} and tagged it according to the features used in the file.
Table~\ref{tab:dts-parse-stats} summarizes the results.
The \textbf{Count} column indicates the number of declaration files that use a
specific feature. The 3rd column shows the corresponding
percentage with respect to all files in the repository. 

We start with a limited scope on the features with the highest number of
occurrences. Thus, we focus our effort on building an
end-to-end solution that generates useful declaration files. We choose
the types \texttt{string}, \texttt{number}, \texttt{boolean}, unions
thereof, arrays, function callbacks, and optional elements are considered
both for function parameters and interface properties. We also
treat aliases and overloaded functions.
Some common types such as \texttt{any}, literals, generics, and index
signatures are not yet implemented. 
Rare types such as tuples and intersections (that
is, types formed using the intersection operator
\lstinline[language=TypeScript]/&/) were not considered. 

\begin{table}[tp]
  \caption{Usage of TypeScript features on DefinitelyTyped}
  \begin{center}
    \begin{tabular}{ |l|c|r| } 
      \hline
      \textbf{Feature} & \textbf{Count} & \multicolumn{1}{|c|}{\textbf{\%}} \\ 
      \hline
      type-string & 5086 & 76.50\% \\
      optional-parameter & 4915 & 73.93\% \\
      type-boolean & 3891 & 58.53\% \\
      type-number & 3699 & 55.64\% \\
      type-void & 3548 & 53.37\% \\
      type-union & 3456 & 51.99\% \\
      type-function & 3286 & 49.43\% \\
      type-array & 3264 & 49.10\% \\
      type-any & 3127 & 47.04\% \\
      type-literals & 1925 & 28.96\% \\
      alias-type & 1899 & 28.56\% \\
      index-signature & 1271 & 19.12\% \\
      generics-function & 1145 & 17.22\% \\
      dot-dot-dot-token & 882 & 13.27\% \\
      call-signature & 817 & 12.29\% \\
      generics-interface & 718 & 10.80\% \\
      type-object & 661 & 9.94\% \\
      type-undefined & 577 & 8.68\% \\
      type-intersection & 431 & 6.48\% \\
      readonly & 373 & 5.61\% \\
      type-tuple & 355 & 5.34\% \\
      generics-class & 265 & 3.99\% \\
      static & 251 & 3.78\% \\
      private & 82 & 1.23\% \\
      public & 43 & 0.65\% \\
      protected & 19 & 0.29\% \\
      \hline
    \end{tabular}
  \end{center}
  \label{tab:dts-parse-stats}
\end{table}

Interfaces are created by exploring \texttt{getField} and \texttt{methodCall} interactions from the run-time information. We
gather the interactions for a specific argument and build the
interface by incrementally adding new properties. Interactions within
the \texttt{followingInteractions} field are recursively traversed,
building a new interface at each level up to a predefined depth. Interfaces that turn out
to be equivalent are merged.

To construct the type signature for a function, we inspect the
primitive types of each argument as well as the interface created from
interactions on the argument in the function body. We do the same for
the return type of the function. This procedure yields a candidate
type for each function call observed during the execution.
For instance, for the \lstinline{globToRegExp} examples shown in
Listing~\ref{code:code-example-extracted-motivating-example} we obtain
the following observed candidate types
\begin{verbatim}
globToRegExp : (glob: string,
                opts: undefined) -> RegExp

globToRegExp : (glob: string, 
                opts: { extended: bool,
                        globstar: undefined,
                        flags: undefined }) -> RegExp
\end{verbatim}
These types are correct, but underapproximate the best possible type
for the package. Our comparison tool (see Section~\ref{sec:dts-compare}) classifies the differences as
solvable, hence we provide one more code example:
\begin{verbatim}
re = globToRegExp("*/www/{*.js,*.html}", {
  flags: "i",
  globstar: true,
});
\end{verbatim}
From this additional example, we obtain the following type:
\begin{verbatim}
globToRegExp : (glob: string, 
                opts: { extended: undefined,
                        globstar: bool,
                        flags: string }) -> RegExp
\end{verbatim}

As the type for the argument \lstinline{glob} and the return type
\lstinline{RegExp} are the same, we collapse the type for
\lstinline{opts} into a union type. The union with the
\lstinline{undefined} type is special and converted into an optional
argument. Moreover, the union of object types proceeds
componentwise. Altogether, we arrive at a single collapsed candidate
type.
\begin{verbatim}
globToRegExp : (glob: string, 
                opts?: { extended?: bool,
                         globstar?: bool,
                         flags?: string }) -> RegExp
\end{verbatim}
We proceed in this way with merging further candidate types until we
reach a set of non-mergable types for the function. If this set
contains more than one candidate type, then we generate an overloaded
definition. For example, the candidate types
\begin{verbatim}
f(string): string
f(number): number
f(boolean): boolean
\end{verbatim}
cannot be merged as the return types are different. They give rise to
an overloaded definition with three alternatives. However, given the candidate types
\begin{verbatim}
g(string): string
g(number): string
g(boolean): boolean
\end{verbatim}
we can merge the first two types on the first argument (because all
remaining arguments and the return types are equal) by collecting \lstinline{string} and
\lstinline{number} into a union type. The resulting set of non-mergable types
has two elements.
\begin{verbatim}
g(string|number): string
g(boolean): boolean
\end{verbatim}

\section{Evaluation}
\label{sec:dts-generate-evaluation}
After generating a declaration file for an \NPM{} package, we need to
evaluate its quality.
Clearly, the quality of the generated
declaration file is heavily dependent on the quality of the examples
provided by the developers. For
example, if there is no code example exploring an execution path that would
access a particular property, then that
property will not appear in the gathered run-time data and
consequently will not appear in the generated declaration file.

As a measure of the quality of a generated file we use its distance to
the declaration file provided for the same module on
DefinitelyTyped. In other words, we use DefinitelyTyped as our ground
truth, because it is the only available reference against we can
compare our results.
To this end, we created \texttt{dts-compare}, a tool to compute the
differences between two TypeScript declaration files.

\begin{figure}[tp]
  \begin{centering}
      {\includegraphics[width=1\linewidth]{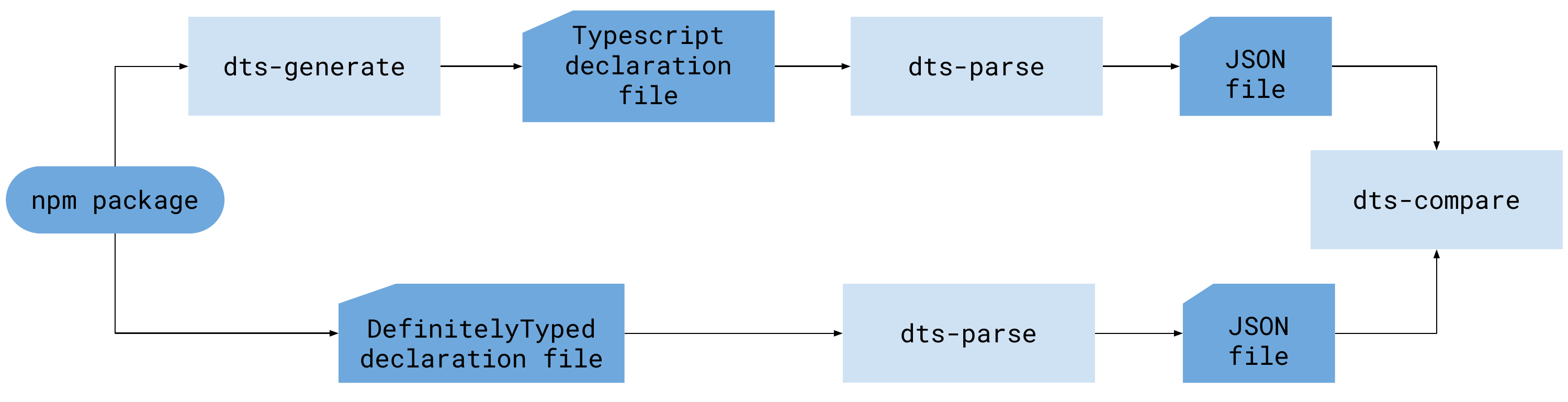}}
      \caption[Evaluation against DefinitelyTyped repository]{{Evaluation of
          generated declaration files
        }
} 
      \label{fig:evaluation-diagram}
  \end{centering}
\end{figure}

For each module, we applied \texttt{dts-compare} to the generated declaration file and the
one uploaded to the DefinitelyTyped repository for the same module as shown in
\figref{fig:evaluation-diagram}. While this approach does not provide an absolute measure
of quality, it gives us at least an indication of the usefulness of \texttt{dts-generate}:
The files on DefinitelyTyped are perceived to be valuable to the community. If the accuracy
of the generated files is comparable with the accuracy of the files on DefinitelyTyped,
then \texttt{dts-generate} is a viable alternative to hand-crafted definition files.

\subsection{dts-parse}
\label{sec:dts-parse}

We are not interested in a textual comparison of declaration files, but in a
comparison of the structures described by the
files. \texttt{dts-parse} receives the file as an input
parameter and returns an AST serialized in JSON.

The first step is to parse the declaration files using the TypeScript
compiler API to build and traverse the 
abstract syntax tree of a TypeScript program
\cite{typescript-compiler-api}. 
This step also performs a sanity check of the generated declaration
files as it rejects files with syntactic or semantic errors.
The output of \texttt{dts-parse} is a structure similar to a symbol table where declared
{interfaces}, {functions}, {classes}, and
{namespaces} are stored separately. Function arguments are
described, identifying complex types like union types or
callbacks. Optional parameters are also identified. For
{classes}, a distinction between constructors and methods
is made. Declaration files are tagged according to their features to identify and filter out declaration files
that contain unimplemented features.


\subsection{dts-compare}
\label{sec:dts-compare}

The compare tool takes two declaration files as input and returns a
JSON file containing the result of the comparison as shown in
\coderef{code:dts-compare-example}. Following the naming convention of unit
testing frameworks, the input files are marked as \lstinline{expected} or
\lstinline{actual}. We use \lstinline{expected} for the DefinitelyTyped
declaration files. 

\begin{lstlisting}[
  caption={Comparing declaration files without differences},
  language=bash,
label=code:dts-compare-example,
  float=tp,captionpos=b
]
$ dts-compare -e expected.d.ts -a actual.d.ts --module-name "my-module"
{
    "module": "my-module",
    "template": "module",
    "differences": [],
    "tags": []
}
\end{lstlisting}

\texttt{dts-compare} applies \texttt{dts-parse} internally to both files to get 
normalized structures that are easy to compare. The result of the comparison is an array
of an abstract entity \texttt{Difference}. 
We classify a difference as \texttt{solvable} if it can be resolved by adding more code
examples. For example, we might add a function invocation with different parameters so that a missing
interface property gets accessed. Otherwise it is classified as  \texttt{unsolvable}. 

We consider the following differences:
\begin{itemize}
  \item \texttt{TemplateDifference}: A difference between the choice of TypeScript
    templates. For example, if the file in DefinitelyTyped is written as a \textbf{module}
    but we generated the file using the \textbf{module-function} template. The comparison
    stops if the templates differ. 
  \item \texttt{ExportAssignmentDifference}: An equality check between the export
    assignments of both files, that is in the expression
    \lstinline{export = XXX}.
  \item \texttt{FunctionMissingDifference}: A function declaration is present in the expected
    file but not in the generated one. This difference applies to functions as well as methods of classes
    and properties of interfaces. 
  \item \texttt{FunctionExtraDifference}: A function is present in the generated declaration file but not in DefinitelyTyped.
  \item \texttt{FunctionOverloadingDifference}: The number of declarations for the same
    function is different. This difference can be due to inexperience of the author of a
    declaration file, who uses, e.g., multiple declarations where a union typed argument
    would be appropriate.
  \item \texttt{ParameterMissingDifference}: A parameter of a function or a property of an interface is not present in the generated file.
  \item \texttt{ParameterExtraDifference}: A parameter of a function or a property of an
    interface is present in the generated file but not in the DefinitelyTyped file. 
  \item \texttt{ParameterTypeDifference}: A parameter of a function or a property of an interface is generated with a different type than in the DefinitelyTyped file. Here we differentiate between \texttt{SolvableDifference} and \texttt{UnsolvableDifference}.
  \begin{itemize}
    \item \texttt{SolvableDifference}: A type difference that can be solved by writing
      further code examples. For example, a basic type that is converted to a union type,
      a function overloading, or a parameter or property that is marked as optional. 
    \item \texttt{UnsolvableDifference}: Any difference not considered as \texttt{SolvableDifference}.
  \end{itemize}
\end{itemize}

Type aliases are expanded before the comparison and thus invisible to
\texttt{dts-compare}: the declaration
\lstinline{type T = string | number; declare function F(a: T);} is equivalent to
\lstinline{declare function} \lstinline{F(a: string | number);}. The same approach applies to literal
interfaces. \texttt{dts-compare} does 
not contemplate differences in function return types, because interface inference of
function return types is limited in the current version of
\texttt{dts-generate}. While the return values are inspected, a proper
interface inference requires further operations on the return value
(cf.\ the explanation of interactions in \secref{sec:run-time-information}).

\section{Results}
\label{sec:results}
We analyzed the generated files focusing on the following aspects:
\begin{itemize}
  \item The quality of the inferred types, interfaces, and module structure using run-time information.
  \item The benefits of using code examples provided by developers as
    a first approximation to execute the libraries. 
  \item The usability of the generated declaration file and whether \texttt{dts-generate} can be used in a proper development environment.
\end{itemize}

\begin{quotation}
  The conducted experiments included tests that consisted of replacing
  a specific type definition from DefinitelyTyped
  \cite{definitely-typed-repository} with the one generated in the
  experiments: TypeScript compilation was successful, the generated
  JavaScript code ran without errors and code intelligence features
  performed by IDEs like code completion worked as expected.
\end{quotation}

Declaration files were generated for existing modules uploaded to the
\NPM{} registry. The DefinitelyTyped repository was used as a
benchmark. Each of the generated files was compared against the
corresponding declaration file uploaded to the repository.

\begin{figure}[tp]
  \centering
  \includegraphics[width=1\linewidth]{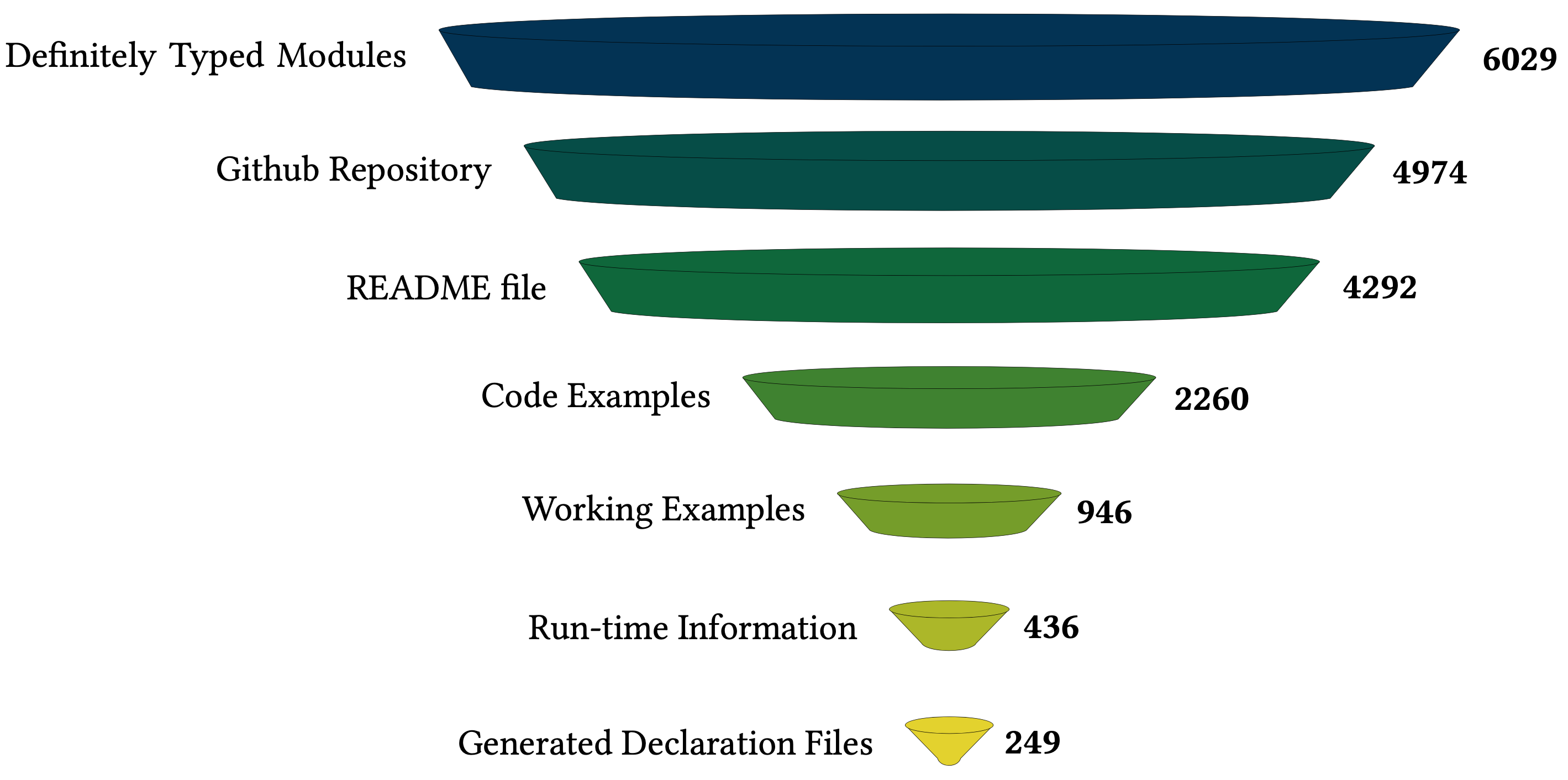}
  \caption[Number of analyzed modules for each stage of the experiment]{
    {Number of analyzed modules at each stage}
  } 
  \label{fig:experiments-overall-funnel}
\end{figure}

Figure \ref{fig:experiments-overall-funnel} shows that a declaration file was generated
for \CountModulesGeneratedDeclarationFile{} modules out of
\CountTotalModulesDefinitelyTyped{} modules on DefinitelyTyped and we identified \CountModulesOnlySolvableDifferences{} modules that have only the
features implemented by \texttt{dts-generate}. We obtained positive results for all of
them. Section~\ref{sec:experiments-evaluation} provides a detailed explanation of the overall
quality of the generated files.
Section~\ref{sec:experiments-declaration-files-generation} presents examples of the generated declaration files for
templates \textbf{module}, \textbf{module-class}, and \textbf{module-function}.

\subsection{Code Examples}
Retrieving the code examples for the JavaScript libraries proved to be
a pragmatic way of driving the type gathering at run time.
\figref{fig:experiments-overall-funnel} gives an overview of the
process to extract code examples. To understand why it was only possible
to obtain working code examples for \CountModulesWithCodeExamples{}
packages, we explain the steps in the process and analyze the losses
in each step.

The process of getting a valid code example for a module is divided in four
stages: 
\begin{itemize}
\item extracting the repository URL;
\item extracting a README file;
\item extracting code examples from README files;
\item executing code examples and discarding failing ones.
\end{itemize}

Next, we describe the results obtained for each step. 

\paragraph*{Repository URL}
The URL of the source repository could be retrieved for only \CountModulesWithRepositoryUrl{}
packages. More than 1000 packages on \NPM{} do not have a repository
entry in their corresponding \texttt{package.json} file. Therefore, the
\texttt{npm view <module> repository.url} command returns no
value. Even important modules like \texttt{ace} provide no repository URL.

\paragraph*{README Files}
\CountModulesWithoutReadmeFile{} packages do not have a README file in their repository, although
the implementation checks for several naming conventions like
\texttt{readme.md} or \texttt{README.md}. 

\paragraph*{Extraction of Code Examples}
In this step, we lose another 50\% of modules! This loss is mainly
explained because some developers do not wrap their code in a block
using the \texttt{javascript} or \texttt{js} tags. As we are
still left with code examples for \CountModulesWithCodeExamples{} modules, we did not look
further into code extraction as this number was considered sufficient for
evaluating the generation of declaration files. 

\paragraph*{Execution of Code Examples}
We executed the remaining \CountModulesWithCodeExamples{} extracted code examples by installing the
required packages and running the code as a \NodeJS{}
application. Unfortunately,
\CountModulesNotWorkingCodeExamples{} modules did not run correctly and had to be
discarded.
The code examples worked for the remaining \CountModulesWorkingCodeExamples{}
modules.
Some failing samples were analyzed and there were mainly
two reasons for the failure: 
\begin{itemize}
\item The code fragment had been properly extracted but the code was
  faulty. It invoked the library in an unsupported
  (obsolete?) way, which lead to a run-time error.
\item The extracted code fragment was not intended to be executed
  and/or it was not even valid JavaScript code. 
\end{itemize}

\paragraph*{Run-time Information}
Run-time information was extracted for only \CountModulesRunTimeInfoExtracted{} out of \CountModulesWorkingCodeExamples{} modules with working code
examples. As explained, to extract the run-time information, the behavior of the code
under analysis was explicitly modified by wrapping the arguments. Furthermore, Jalangi's
instrumentation itself caused some executions to fail, because the modules contained
JavaScript features that are not supported by Jalangi. As a result, run-time information
could not be extracted for \CountModulesRunTimeInfoCouldNotBeExtracted{} modules. An instrumentation without user-defined analysis
modules was not applied, so it was not possible to determine which modules were failing
only because of Jalangi's own limitations. 

\paragraph*{Generated Declaration Files}
A declaration file was generated for \CountModulesGeneratedDeclarationFile{} out of \CountModulesRunTimeInfoExtracted{} modules. Despite the correct execution
of the instrumented code examples, the extracted code for
\CountModulesDeclarationFileCouldNotBeGenerated{} modules did 
not exercise the library sufficiently. Hence, the collected run-time information was not enough
for generating a declaration file. 

\subsection{Declaration Files Generation}
\label{sec:experiments-declaration-files-generation}

This section exhibits an example of the \CountModulesGeneratedDeclarationFile{} generated
declaration files. \figref{fig:experiments-results-module-steamid} exposes a simplified example for the
\textbf{module-class} template where we highlight the most important features.
We expose the results for the remaining templates \textbf{module} and \textbf{module-function} in
\secref{sec:appendix-results-templates}.
The left side of the figure shows
the generated declaration file with x
\lstinline{dts-generate}, the right side shows the corresponding file
in the DefinitelyTyped repository. 

\begin{figure*}[tp]
  \centering
  \begin{subfigure}[t]{0.48\linewidth}
    \begin{lstlisting}[language=TypeScript,numbers=none]
export = SteamID;
declare class SteamID {
    // ...
    isValid(): boolean;
    isGroupChat(): boolean;
    isLobby(): boolean;
    getSteam2RenderedID(newerFormat?: boolean): string;
    // ...
}
declare namespace SteamID {
    export function fromIndividualAccountID(accountid: string | number): SteamID;
}  
    \end{lstlisting}
    \caption{steamid/index.d.ts - Generated (simplified)}
  \end{subfigure}
  \hfill
  \begin{subfigure}[t]{0.48\linewidth}
    \begin{lstlisting}[language=TypeScript,numbers=none]
// ...
type getSteam2RenderedID = (newerFormat?: boolean) => string;

declare class SteamID {
    // ...
    isValid(): boolean;
    // ...
    isGroupChat(): boolean;
    // ...
    isLobby(): boolean;
    getSteam2RenderedID: getSteam2RenderedID;
    // ...
}

declare namespace SteamID {
    // ...
    function fromIndividualAccountID(accountid: number | string): SteamID;
}

export = SteamID;
    \end{lstlisting}
    \caption{steamid/index.d.ts - DT (simplified)}
  \end{subfigure}

  \caption{Results for \textbf{module-class} template for module \texttt{steamid}}
  \label{fig:experiments-results-module-steamid}
\end{figure*}

There are no relevant differences between the files. 
Class methods and functions are correctly detected; input and output types are
accurately inferred. We extract the parameter names from the variable names of the JavaScript
code. For the \texttt{getSteam2RenderedID} method we correctly generate a signature
that matches the type alias chosen by the author of the library. Finally, the method \\
\texttt{fromIndividualAccountID} is correctly located in the namespace section.

It is worth mentioning that for some libraries the declaration file in DefinitelyTyped was
not correct. For module \texttt{glob-base} we detected that a parameter was
incorrectly marked as optional. For module \texttt{smart-truncate} we discovered that the author
incorrectly used the \textbf{module} template.
We provide a more detailed explanation in \secref{sec:errors-definitely-typed}.

\subsection{Evaluation}
\label{sec:experiments-evaluation}
We analyzed \CountModulesOnlySolvableDifferences{} of the \CountModulesGeneratedDeclarationFile{} generated files. The remaining 174 files contained at least one
of the unimplemented TypeScript features so that they could not be
considered.

The code examples have a direct influence on the quality of the generated declaration file. We introduced the
concept of \texttt{solvable} and \texttt{unsolvable} differences in
\secref{sec:dts-compare}. Modules containing only \texttt{solvable} differences were
considered as positive results. All of the analyzed files contained only \texttt{solvable} differences.
We manually completed the
examples for 10 of those modules. Applying \texttt{dts-generate} with the completed
examples generates declaration files which are equivalent to the files from DefinitelyTyped.
\secref{sec:appendix-manually-completed-examples} exhibits the manually added code examples
for module \texttt{github-url-to-object}.

\section{Discussion}
\label{sec:discussion}

How can we improve further? Given that JavaScript developers are incentivized to develop good
examples rather than writing ``boring'' type declarations, how can we capitalize on that attitude?
The main issue is that any example-driven algorithm can only come up with an
under-approximation of the intended meaning, so some form of help is required when
generalization is desired.

We believe that  JavaScript developers can be taught to write their
code examples in such a style that high-quality TypeScript signatures
can be extracted from them without human intervention. 
In the subsequent paragraphs we survey some approaches that rely on
developers using special strings and other constants in their code examples.

\paragraph*{Literal types vs string}
Many JavaScript libraries use literal strings to select
options and configurations. To be concrete, 1925 DT packages rely on
literal types (see Table~\ref{tab:dts-parse-stats}). For example, the bonjour
package\footnote{\url{https://www.npmjs.com/package/@types/bonjour}}
relies on strings to control events and indicate protocols in (different) interfaces:
\begin{lstlisting}[language=TypeScript]
    removeAllListeners(event?: 'up' | 'down'): this;
    protocol?: 'udp'|'tcp';
\end{lstlisting}
The carlo
package\footnote{\url{https://www.npmjs.com/package/@types/carlo}}
uses literals to distinguish different events.
\begin{lstlisting}[language=TypeScript]
export type AppEvent = 'exit' | 'window';
\end{lstlisting}

For our framework, it is hard to distinguish a string that is meant literally from a
string that just serves as an example. An automatic solution might collect
typical values of literal strings and generate the corresponding literal types (or
unions thereof) when the arguments used in the example are all typical. Alternatively, one
might always generate literal types until the number of different
examples passes some threshold, in which case the generator
generalizes to type 
\lstinline/string/. Both alternatives come with the problem that they
create a local over-approximation, which makes it harder to
qualify the output of the generator. A further alternative that does
not suffer from this drawback would be to communicate to the developer a set
of typical example strings, \emph{marker strings}, that the generator
always generalizes to the \lstinline/string/ type.

\paragraph*{Any}
Suppose the programmer supplies examples that use \lstinline/number/, \lstinline/bool/, and
\lstinline/string/ at the same argument position. In this case, our
tool calculates the union type \lstinline/number|bool|string/. But what if the programmer
wants to advertise the type \lstinline/any/ for an argument? This intent is hard to
communicate with a finite number of examples.

We propose that the programmer relies on marker strings like \lstinline/"any"/ to indicate
the \lstinline/any/ type for an argument position.

\paragraph*{Further types}
Similar approaches could be conceived for indexed types, polymorphic types, dependent
types, etc. At present we believe incorporating these
features in a tool to be of limited value because they are not widely used in DefinitelyTyped (see
Table~\ref{tab:dts-parse-stats}). 

\section{Related Work}
\label{sec:related-work}
\paragraph*{Dynamic analysis techniques} Trace Typing
\cite{DBLP:conf/ecoop/AndreasenGCSTS16} is a framework for evaluating
retrofitted type systems. It involves gathering traces of JavaScript program
executions, just like in our system. However, the goal of the type
systems that they study is quite different from ours. They are
interested in type information to improve run-time performance and
traditional compilation, giving information about object layout and
tag tests. Contrary to their approach, we do not just observe the
types of arguments and return values. In addition, we also trace
property accesses which translate to requirements on objects. We
have no generalization step to infer polymorphic types.

Hummingbird \cite{DBLP:conf/pldi/RenF16} is a system for type checking
Ruby programs at run time in the presence of
metaprogramming. Metaprograms additionally generate
type annotations for the generated code. Once a generated method is
called, the method's body is statically checked against the actual
argument types as well as the generated annotations. This system does
not involve run-time tracing.

Rubydust \cite{DBLP:conf/popl/AnCFH11} implements dynamic
constraint-based type inference for Ruby. Rubydust runs an
instrumented Ruby program that gathers subtyping constraints for
parameters and return types of methods as well as for fields, which is
quite different from the interaction that we collect. After
completion, the solution of these constraints is proposed as a typing
for the program. These types are sound if the runs cover sufficiently
many paths in the program. Rubydust is reasonably effective because
Ruby's type system is nominal. For JavaScript, this approach is much
less effective as (run-time) typing is structural
\cite{DBLP:conf/sfp/NausT16}.

JSTrace \cite{saftoiu21:_jstrac} performs dynamic type inference
for \\ JavaScript by gathering information at run time. It inspects
run-time types of arguments and return values of suitably wrapped
functions. While the resulting information is treated similarly as in
our work, we gather interactions from property
accesses and method calls and derive further typing constraints from
those. Moreover, JSTrace performs an (expensive) deep inspection of
the types of objects and arrays whereas our inspection is shallow and
thus more efficient.

None of the works tackles the question how to obtain a sufficient
sample of run-time information. One part of our contribution is to
systematically uncover the README files as a driver for running the
JavaScript code under inspection.

\paragraph{Data Type Inference}
Petricek and coworkers describe a mechanism to infer a datatype from
an example input in JSON, XML, or CSV format
\cite{DBLP:conf/pldi/PetricekGS16}. Their implementation relies on
type providers, a metaprogramming facility of F\#. The type provider reads the
example input at compile time and translates it into a static
type. The approach is purely data-centric---the shape of the sample
data determines what is legal in the program. Our approach
analyzes the program's behavior on sample data and generalizes these
observations to a type.

\paragraph*{dts-gen}
TypeScript comes with \texttt{dts-gen}, a tool that creates declaration files for
JavaScript libraries \cite{dts-gen}. Its documentation states that the result is 
intended to be used as a starting point for the development of a declaration file. The outcome
needs to be refined afterwards by the developer. 

The tool analyzes the shape of the objects at run time after initialization without
executing the library. This results in most parameters and results being inferred as
\lstinline[language={}]{any}.

In contrast, our tool \texttt{dts-generate} is intended to generate declaration files
that are ready to be uploaded to DefinitelyTyped without further manual intervention. Any
amount of manual work that a developer needs to do on a declaration file after updating
JavaScript code increases the risk of discrepancies between the declaration file
and the implementation.







\paragraph*{TSInfer \& TSEvolve}
TSInfer and TSEvolve are presented as part of TSTools
\cite{DBLP:conf/fase/KristensenM17}. Both tools are the continuation of TSCheck
\cite{DBLP:conf/oopsla/FeldthausM14}, a tool for detecting mismatches between a
declaration file and the implementation of the module. 

TSInfer proceeds in a similar way than TSCheck. It initializes the library in a browser
and records a snapshot of the resulting state.  Then it performs a lightweight static
analysis on all the functions and objects stored in the snapshot. 

The abstraction and the constraints introduced as part of the static analysis tools
for inferring the types have room for improvement. A run-time based approach like the one
presented in our work will provide more accurate information, thus generating more precise
declaration files.

TSInfer faces the problem
of including internal methods and private properties from the snapshot in the declaration file. Run-time
information would have shown that the developer has no intention of exposing these
methods. 

Moreover, TSEvolve performs a differential analysis on the changes made to a JavaScript
library to determine intentional discrepancies between the declaration files of two
consecutive versions. However, a differential analysis may not be needed. If the
developer's intention were accurately represented by the extracted code, then the generated
declaration file would already describe the newer version of a library without the need of
a differential analysis.

\paragraph*{TSTest}
TSTest is a tool that checks for mismatches between a declaration file and the JavaScript
implementation of the module \cite{DBLP:journals/pacmpl/KristensenM17}. It applies feedback-directed
random testing for generating type test scripts. These scripts execute the library with
random arguments with the typings from the declarations file and check whether the output
matches the prescribed type. TSTest thus provides concrete counterexamples if it detects mismatches.



TSTest could be used to extend our tool with a feedback loop. If TSTest detects a problem
with a declaration file generated by \texttt{dts-generate}, then we would add the
resulting counterexamples to the example code and restart the generation process. 

\section{Conclusions and Future Work}
\label{sec:conclusion}
We have presented \texttt{dts-generate}, a tool for generating a TypeScript declaration
file for a specific JavaScript library. It downloads the developer's
code examples from the library's repository. It uses these examples to
execute the library and gather data flow and type information. The tool generates a TypeScript declaration
file based on the information gathered at run time.


Building an end-to-end solution for the generation of TypeScript declaration files was
prioritized over type inference accuracy. Hence, types were taken over from the
values at run time. Developers express their intent how a library should be used with
example code in the documentation. 
Obtaining the types from the code examples extracted from the repositories proved to be a
pragmatic and effective approximation, enabling to work on specific aspects regarding the
TypeScript declaration file generation itself.

We built a mechanism to automatically create declaration files for potentially every
module uploaded to DefinitelyTyped. We managed to generate declaration files for
\CountModulesGeneratedDeclarationFile{} modules. We compared the results against the corresponding 
files uploaded to
DefinitelyTyped by creating \texttt{dts-parse}, a TypeScript declaration files parser and
\texttt{dts-compare}, a comparator.

Return types could be defined more accurately by analyzing the interactions tracked on
them. We consider that enhancing the run-time information by executing tests already
present in the codebase is an approach worth following as future work.



\clearpage
\appendix
\section{Results}
\subsection{Templates}
\label{sec:appendix-results-templates}
\subsubsection{module-function}
\figref{fig:experiments-results-module-function} shows the generated
declaration files for simple modules like \texttt{abs},
\texttt{dirname-regex}, and \texttt{escape-html}. All of them were
generated using the \textbf{module-function} template. 

The different names in the export assignment by modules \\
\texttt{dirname-regex} and
\texttt{escape-html} do not matter because the module can be named arbitrarily when it is imported: in
\lstinline[language=TypeScript]/import Abs = require('abs');/ the
identifier \lstinline[language=TypeScript]/Abs/ can be chosen by the
programmer. We generate the name in the export 
assignment by transforming the module name into camel case form, following TypeScript
guidelines.

A difference in the names of the function parameters does not affect the correctness of
the declaration file.
Furthermore, neither the order of declarations nor unused namespaces
(as in module \texttt{escape-html}) matter.

\begin{figure*}[tp]
  \centering
  \begin{subfigure}[t]{0.48\linewidth}
    \begin{lstlisting}[language=TypeScript,numbers=none]
export = Abs;

declare function Abs(input: string): string;
    \end{lstlisting}
    \caption{abs/index.d.ts - Generated}
  \end{subfigure}
  \hfill
  \begin{subfigure}[t]{0.48\linewidth}
    \begin{lstlisting}[language=TypeScript,numbers=none]
declare function Abs(input: string): string;
export = Abs;
    \end{lstlisting}
    \caption{abs/index.d.ts - DT}
  \end{subfigure}

  \begin{subfigure}[t]{0.48\linewidth}
      \begin{lstlisting}[language=TypeScript,numbers=none]
export = DirnameRegex;

declare function DirnameRegex(): RegExp;
      \end{lstlisting}
      \caption{dirname-regex/index.d.ts - Generated}
    \end{subfigure}
    \hfill
    \begin{subfigure}[t]{0.48\linewidth}
      \begin{lstlisting}[language=TypeScript,numbers=none]
export = dirnameRegex;

declare function dirnameRegex(): RegExp;
      \end{lstlisting}
      \caption{dirname-regex/index.d.ts - DT}
    \end{subfigure}

    \begin{subfigure}[t]{0.48\linewidth}
      \begin{lstlisting}[language=TypeScript,numbers=none]
export = EscapeHtml;

declare function EscapeHtml(string: string): string;

      \end{lstlisting}
      \caption{escape-html/index.d.ts - Generated}
    \end{subfigure}
    \hfill
    \begin{subfigure}[t]{0.48\linewidth}
      \begin{lstlisting}[language=TypeScript,numbers=none]
declare function escapeHTML(text: string): string;
declare namespace escapeHTML { }

export = escapeHTML;
      \end{lstlisting}
      \caption{escape-html/index.d.ts - DT}
    \end{subfigure}

  \caption{Results for \textbf{module-function} template}
  \label{fig:experiments-results-module-function}
\end{figure*}

\subsubsection{module}
\figref{fig:experiments-results-module-is-uuid} contains an example
for the \textbf{module} template, the declaration for the \texttt{is-uuid} module.
The generated file only contain methods that were executed by the extracted
examples. All invoked functions are correctly detected; the additional
functions declared on DefinitelyTyped are not used in the example
code.
It would be easy to generate a perfectly matching declaration file by
adding two examples using functions \texttt{nil} and
\texttt{anyNonNil}.

\begin{figure*}[tp]
  \centering
  \begin{subfigure}{0.48\linewidth}
    \begin{lstlisting}[language=TypeScript,numbers=none]
export function v1(str: string): boolean;
export function v2(str: string): boolean;
export function v3(str: string): boolean;
export function v4(str: string): boolean;
export function v5(str: string): boolean;
    \end{lstlisting}
    \caption{is-uuid/index.d.ts - Generated}
  \end{subfigure}
  \hfill
  \begin{subfigure}{0.48\linewidth}
    \begin{lstlisting}[language=TypeScript,numbers=none]
export function v1(value: string): boolean;
export function v2(value: string): boolean;
export function v3(value: string): boolean;
export function v4(value: string): boolean;
export function v5(value: string): boolean;
export function nil(value: string): boolean;
export function anyNonNil(value: string): boolean;
    \end{lstlisting}
    \caption{is-uuid/index.d.ts - DefinitelyTyped}
  \end{subfigure}

  \caption{Results for \textbf{module} \texttt{is-uuid}}
  \label{fig:experiments-results-module-is-uuid}
\end{figure*}

\subsection{Manually completed examples}
\label{sec:appendix-manually-completed-examples}
\figref{fig:experiments-results-manually-completed-examples} exhibits the
 manually expanded code examples for module \texttt{github-url-to-object}.

\begin{figure*}[t]
  \centering
  \begin{subfigure}[t]{0.45\linewidth}
    \begin{lstlisting}[language=TypeScript]
declare namespace gh {
  interface Options {
      enterprise?: boolean;
  }
  interface Result {
      ...
  }
}

declare function gh(url: string | {url: string}, options?: gh.Options): gh.Result | null;

export = gh;    
    \end{lstlisting}
    \caption{github-url-to-object/index.d.ts - DefinitelyTyped. Properties of interface \texttt{Result} were ignored for readability, since return types were not considered.}
  \end{subfigure}
  \hfill
  \begin{subfigure}[t]{0.45\linewidth}
    \begin{lstlisting}[language=TypeScript]
export = GithubUrlToObject;

declare function GithubUrlToObject(repoUrl: string | GithubUrlToObject.I__repoUrl, opts?: GithubUrlToObject.I__opts): object | null;
declare namespace GithubUrlToObject {
  export interface I__repoUrl {
    'url'?: string;
  }

  export interface I__opts {
    'enterprise'?: boolean;
  }

}
    \end{lstlisting}
    \caption{github-url-to-object/index.d.ts - Generated with manually expanded code example}
  \end{subfigure}

  \begin{subfigure}{0.80\linewidth}
    \begin{lstlisting}[language=JavaScript]
var gh = require('github-url-to-object')

gh('github:monkey/business');
gh('https://github.com/monkey/business');
gh('https://github.com/monkey/business/tree/master');
gh('https://github.com/monkey/business/tree/master/nested/file.js');
gh('https://github.com/monkey/business.git');
gh('http://github.com/monkey/business');
gh('git://github.com/monkey/business.git');

// Manually added:
gh('git+https://githuuub.com/monkey/business.git', {});
gh('git+https://githuuub.com/monkey/business.git', {enterprise: true});
gh({url: 'git://github.com/monkey/business.git'});
gh('this is not a proper url');
gh({url: 'this is not a proper url'});
    \end{lstlisting}
    \caption{Code example for module \texttt{github-url-to-object}.}
    \end{subfigure}
  \caption{Manually expanded code example for
    \texttt{github-url-to-object} to match DefinitelyTyped declaration
    file. The literal interface in DefinitelyTyped is replaced by a
    declared interface. Return types are not considered.} 
  \label{fig:experiments-results-manually-completed-examples}
\end{figure*}

\subsection{Errors in DefinitelyTyped}
\label{sec:errors-definitely-typed}

\begin{lstlisting}[
  caption={Code example for \texttt{smart-truncate} module exporting a function},
  label=code:experiments-results-module-smart-truncate,
  language=JavaScript,captionpos=b,float,numbers=none
  ]
  var smartTruncate = require('smart-truncate');
  
  var string = 'To iterate is human, to recurse divine.';
  
  // Append an ellipsis at the end of the truncated string.
  var truncated = smartTruncate(string, 15);
\end{lstlisting}

For module \texttt{glob-base} the parameter
\texttt{basePath} is declared as optional in DefinitelyTyped. However, when invoking the
function without the \texttt{basePath} parameter, an error is thrown at run time. The file generated by
\texttt{dts-generate} does not mark this parameter as optional, as shown in
\figref{fig:experiments-results-module-glob-base}. Function return type interfaces are
not inferred by \texttt{dts-generate}.

We also
discovered errors in the selected template in DefinitelyTyped. The module
\texttt{smart-truncate} in DefinitelyTyped uses the \textbf{module} template, but
\texttt{dts-generate} generates a file using the \textbf{module-function}
template. \coderef{code:experiments-results-module-smart-truncate} shows that the module is indeed
exported as a function, as inferred by \texttt{dts-generate}.

\begin{figure*}[tp]
  \centering
  \begin{subfigure}[t]{0.48\linewidth}
    \begin{lstlisting}[language=TypeScript,numbers=none]
'use strict';

// ...

module.exports = function globBase(pattern) {
if (typeof pattern !== 'string') {
  throw new TypeError('glob-base expects a string.');
}

// ...
} 
    \end{lstlisting}
    \caption{glob-base/index.js - Excerpt of JS implementation
      that throws an error when input parameter is not a string} 
    \label{fig:subfloat-globbase-js-implementation}
  \end{subfigure}
  \hfill
  \begin{subfigure}[t]{0.48\linewidth}
    \begin{lstlisting}[language=TypeScript,numbers=none]
declare function globbase(basePath?: string): globbase.GlobBaseResult;

declare namespace globbase {
  interface GlobBaseResult {
      base: string;
      isGlob: boolean;
      glob: string;
  }
}

export = globbase;
    \end{lstlisting}
    \caption{glob-base/index.d.ts - DefinitelyTyped}
  \end{subfigure}

\begin{subfigure}[t]{0.48\linewidth}
    \begin{lstlisting}[language=JavaScript,numbers=none]
var globBase = require('glob-base');

globBase();

// Throws: TypeError: glob-base expects a string.
    \end{lstlisting}
    \caption{Code example invoking the exported function without the first parameter, as specified in DefinitelyTyped}
  \end{subfigure}
  \hfill
  \begin{subfigure}[t]{0.48\linewidth}
    \begin{lstlisting}[language=TypeScript,numbers=none]
export = GlobBase;

declare function GlobBase(pattern: string): object;
    \end{lstlisting}
    \caption{glob-base/index.d.ts - Generated}
  \end{subfigure}
  
  \caption{Incorrect optional parameter for module \texttt{glob-base}}
  \label{fig:experiments-results-module-glob-base}
\end{figure*}

\clearpage

\bibliographystyle{ACM-Reference-Format}
\bibliography{main}

\end{document}